# Process Diagrams

Sheelagh Carpendale, Computing Science, Simon Fraser University

**Abstract:** This paper is simply a collection of process diagrams for further use and reference. These diagrams about different approaches to research.

## Introduction

These are just a few diagrams that I have found useful for explaining some research processes. Figure 1 is a largely verbal explanation of the scientific experimental research process where one carefully develops a hypothesis, selects appropriate independent variable, works hard at eliminating complexity to arrive a result that can be declared with a reasonable degree of certainty. Normally one likes to have a 95% confidence rating. Figure 2 shows the expansive process of observation for design. Here one works towards a frank and reflective assessment of ones own starting position to be conscious, and thus less influenced by, one's own biases and beginnings. New insight and understanding bubbles up for the richness and complexity of reality.

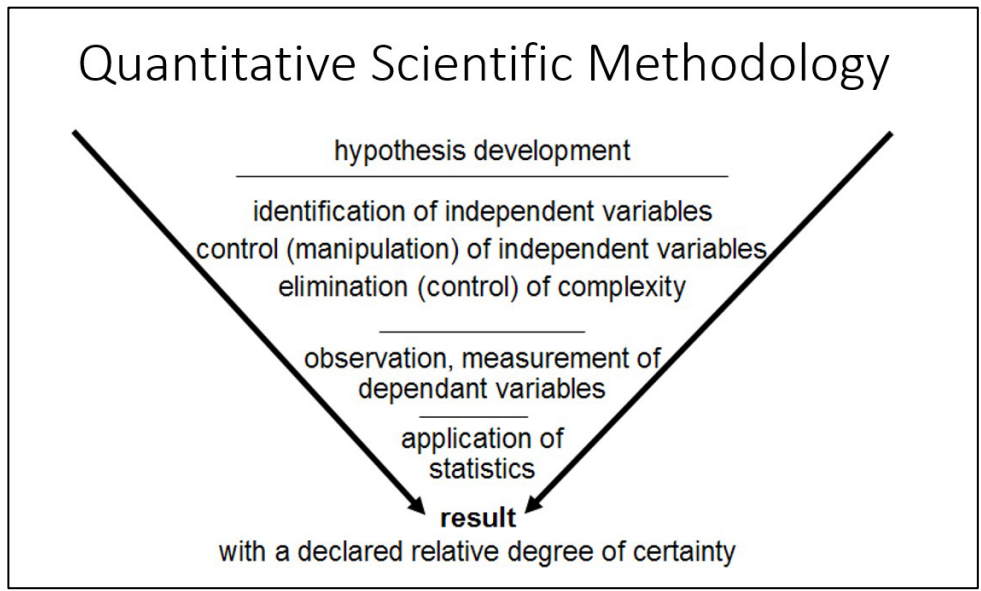

*Figure 1: this is a diagram of the basic steps in the scientific experimental method*

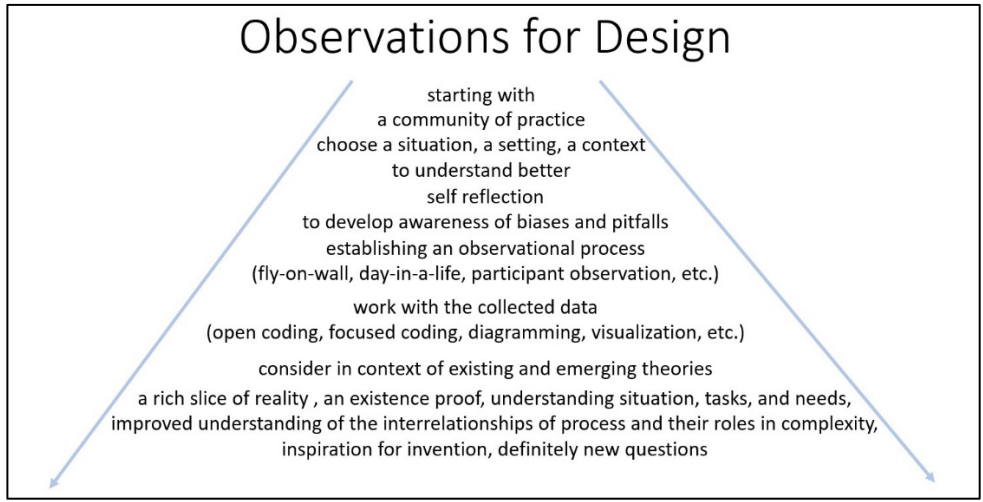

*Figure 2: Conducting careful observations, doing qualitative analysis, reaching sufficient understanding to see possibilities for better solutions is a design approach I particularly like.*

Figure 3 is two diagrams in one. The top half shows the iterative process of DSM (Design Study Methodology [4]), where one often returns to previous steps to retrace and repeat steps to gradually work towards a mutually satisfying end result. The lower half shows the visualization reference diagram [1] with links to illustrate how these two diagrams might relate to each other.

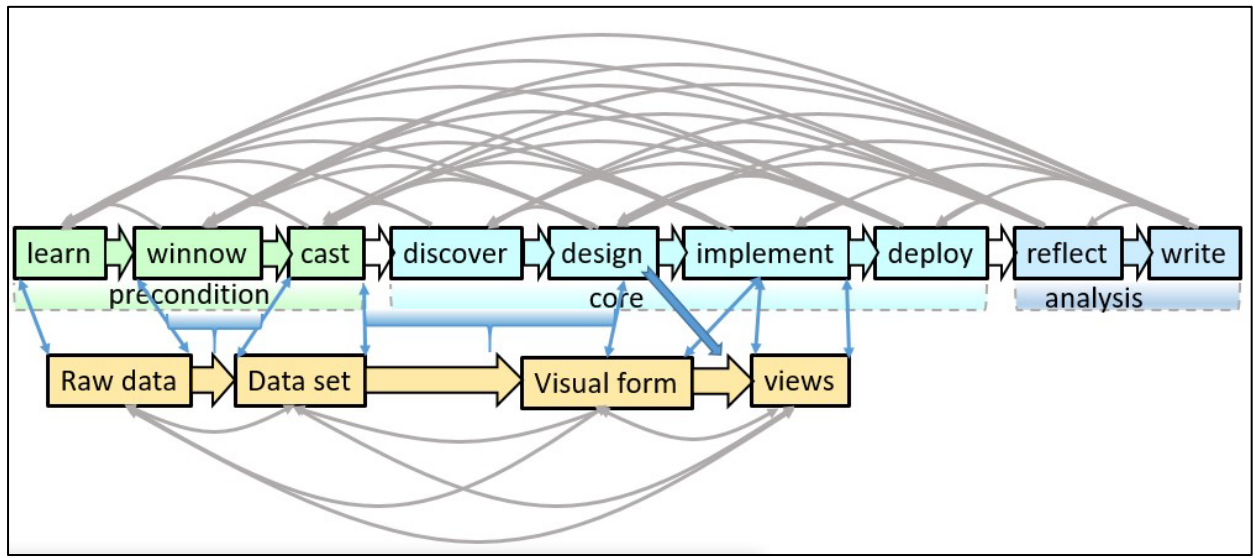

*Figure 3: this diagram shows the relationship between the diagram of the Design Space Methodology [4] and the visualization reference diagram [1]. Notice how the last three steps in the design space methodology (deploy, reflect, and write) are not part of the process in the reference diagram.*

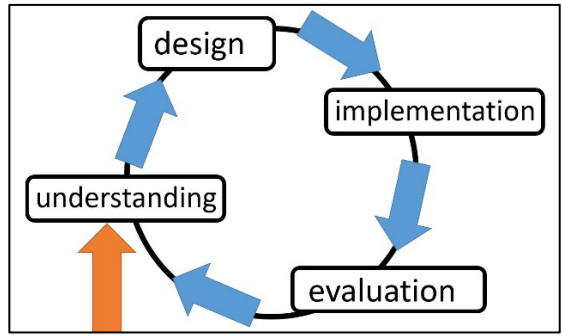

*Figure 4: These four processes design, implementation, evaluation, and analysis to develop understanding, are all important steps in tool making that are usually done in this order. However, the process can be started from any of the four points.*

Figure 4 is an illustration of the often discussed cyclical nature of tool creation. Figure 5 is simply and re-drawing of the used in Lee et al. [2] article on creating data driven stories.

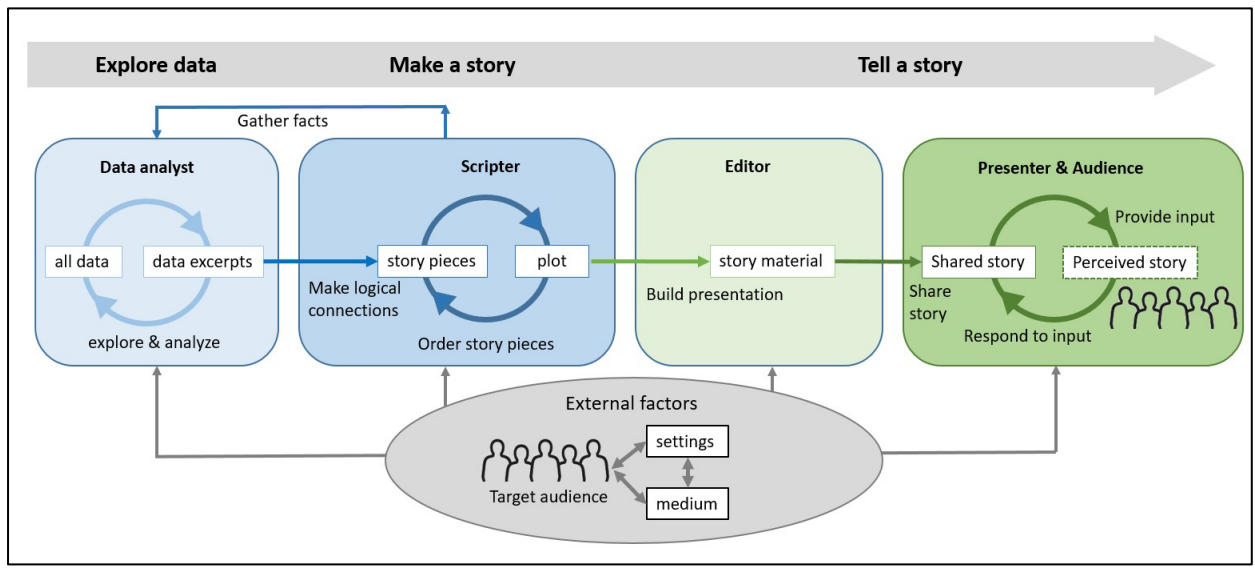

*Figure 5: This diagram is a re-drawing of the diagram in Lee et al. [2] More than Telling a Story.*

Figure 6 only makes a small point about Autographic visualization. See the diagrams in Offenhuber [3] for fuller explanations. Figure 6 shows how that once a phenomenon is selected it is not always a simple matter to be able to read the underlying data. There can be considerable design thinking and design work to create a way in which it is easier for us a human to be able to read the data in the phenomena.

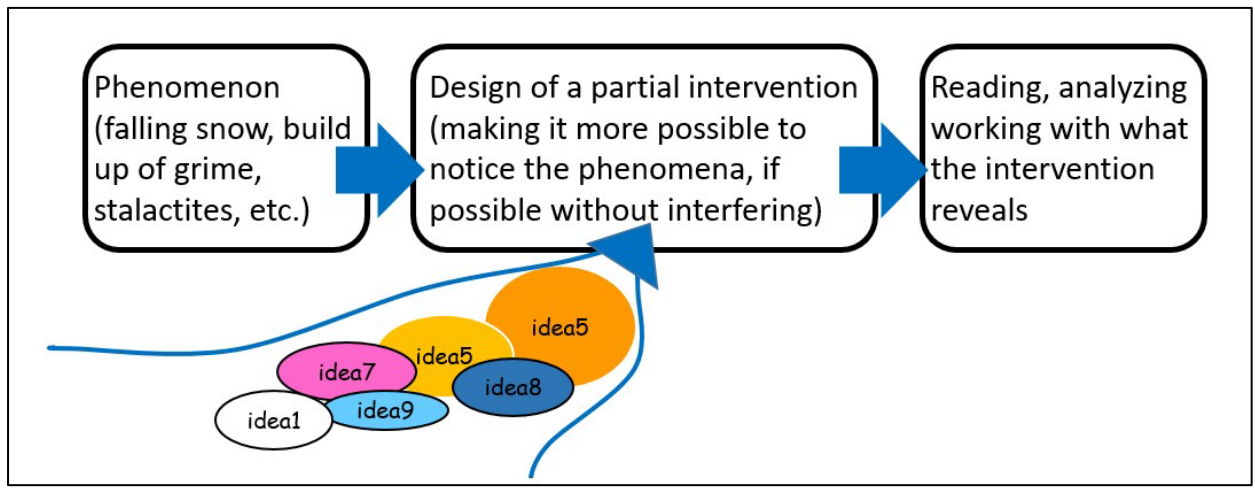

*Figure 6: This diagram does not explain the full Autographic visualization [3] process. It simply shows that a lot of creative design work happens in the middle step when trying to work out how to make the phenomena understandable.*

**References**

1. Card, Mackinlay. *Readings in information visualization: using vision to think*. Morgan Kaufmann, 1999.
2. Lee, Bongshin, Nathalie Henry Riche, Petra Isenberg, and Sheelagh Carpendale. "More than telling a story: Transforming data into visually shared stories." *IEEE computer graphics and applications* 35, no. 5 (2015): 84-90.
3. Offenhuber, Dietmar. "Data by proxy—material traces as autographic visualizations." *IEEE transactions on visualization and computer graphics* 26, no. 1 (2019): 98-108.
4. Sedlmair, Michael, Miriah Meyer, and Tamara Munzner. "Design study methodology: Reflections from the trenches and the stacks." *IEEE transactions on visualization and computer graphics* 18, no. 12 (2012): 2431-2440.